\documentclass[12pt,letterpaper]{article}
\usepackage[a4paper, total={7in, 10in}]{geometry}

\usepackage{graphicx}
\usepackage{helvet}
\usepackage{authblk}
\usepackage{hyperref}
\usepackage{amsmath} 
\usepackage{amssymb} 
\usepackage{orcidlink} 
\usepackage[super,comma,sort&compress]  
   {natbib}
\usepackage[right]{lineno} \linenumbers
\usepackage[title]{appendix}%

\usepackage[utf8]{inputenc} 
\usepackage[T1]{fontenc}    
\usepackage{url}            
\usepackage{booktabs}       
\usepackage{amsfonts}       
\usepackage{nicefrac}       
\usepackage{microtype}      
\usepackage{xcolor}         
\usepackage{multirow}
\usepackage{threeparttable}
\usepackage{amsmath}
\usepackage{subcaption}
\usepackage{makecell}
\usepackage{placeins}
\usepackage[utf8]{inputenc} 
\usepackage{booktabs}      
\usepackage{siunitx}       
\usepackage{caption}       
\usepackage{tabularx}
\usepackage{listings}    
\usepackage{pifont}
\usepackage{array}
\usepackage{makecell}
\theadset{\bfseries}

\makeatletter
\renewcommand{\maketitle}{\bgroup\setlength{\parindent}{0pt}
\begin{flushleft}
  \textbf{\@title}
  
  \@author
\end{flushleft}\egroup}
\makeatother

\title{Artificial Intelligence-derived Cardiotocography Age as a Digital Biomarker for Predicting Future Adverse Pregnancy Outcomes}

\author[1,2]{Jinshuai Gu}
\author[2]{Zenghui Lin}
\author[2,6]{Jingying Ma}
\author[7]{Jingyu Wang}
\author[3]{Linyan Zhang}
\author[3]{Rui Bai}
\author[3]{Zelin Tu}
\author[3]{Youyou Jiang}
\author[2]{Donglin Xie}
\author[1,4,*]{Yuxi Zhou}
\author[3,*]{Guoli Liu}
\author[2,5,*]{Shenda Hong}

\affil[1]{Department of Computer Science, Tianjin University of Technology, Tianjin, China}
\affil[2]{National Institute of Health Data Science, Peking University, Beijing, China}
\affil[3]{Department of Obstetrics and Gynecology, Peking University People’s Hospital, Beijing, China}
\affil[4]{DCST, BNRist, RIIT, Institute of Internet Industry, Tsinghua University, Beijing, China}
\affil[5]{Institute for Artificial Intelligence, Peking University, Beijing, China}
\affil[6]{Saw Swee Hock School of Public Health, National University of Singapore, Singapore, Singapore}
\affil[7]{Department of Respiratory and Critical Care Medicine, Binzhou Medical University Hospital, Binzhou, China}

\affil[*]{Correrpondence: hongshenda@pku.edu.cn, joy\_yuxi@pku.edu.cn, guoleeliu@163.com}

\begin{document}

\maketitle

\section*{ABSTRACT}
    \textbf{Background and Objective:} Cardiotocography (CTG) is a low-cost, non-invasive fetal health assessment technique used globally, especially in underdeveloped countries. However, it is currently mainly used to identify the fetus’s current status (e.g., fetal acidosis or hypoxia), and the potential of CTG in predicting future adverse pregnancy outcomes has not been fully explored. We aim to develop an AI-based model that predicts biological age from CTG time series (named CTGage), then calculate the age gap between CTGage and actual age (named CTGage-gap), and use this gap as a new digital biomarker for future adverse pregnancy outcomes.

    \textbf{Methods:} The CTGage model is developed using 61,140 records from 11,385 pregnant women, collected at Peking University People's Hospital (Beijing, China) between January 1, 2018, and December 31, 2022. For model training, a structurally designed 1D convolutional neural network (named Net1D) is used, incorporating distribution-aligned augmented regression technology. The CTGage-gap is categorized into five groups: < –21 days (underestimation group), –21 to –7 days, –7 to 7 days (normal group), 7 to 21 days, and > 21 days (overestimation group). We further defined the underestimation group and overestimation group together as the high-risk group. We then compare the incidence of adverse outcomes (including premature infants, low birth weight infants, neonatal asphyxia, fetal distress, fetal malformations and congenital heart disease) and maternal diseases (including gestational diabetes mellitus, anaemia, maternal congenital disease, umbilical cord problems and placental lesions) across these groups.

    \textbf{Results:} The average absolute error of the CTGage model is 10.91 days. When comparing the overestimation group with the normal group, premature infants incidence is 5.33\% vs. 1.42\% (p < 0.05) and gestational diabetes mellitus (GDM) incidence is 31.93\% vs. 20.86\% (p < 0.05). When comparing the underestimation group with the normal group, low birth weight incidence is 0.17\% vs. 0.15\% (p < 0.05) and anaemia incidence is 37.51\% vs. 34.74\% (p < 0.05). 

    \textbf{Conclusion:} Artificial intelligence-derived CTGage can predict the future risk of adverse pregnancy outcomes and holds potential as a novel, non-invasive, and easily accessible digital biomarker.


\section*{KEYWORDS}
    Cardiotocography, Deep Learning, Biological Age, Adverse Pregnancy Outcomes, Digital Biomarkers

\section*{INTRODUCTION}
    Abnormal fetal growth and development increase the risk of adverse pregnancy outcomes \cite{lubrano2022perinatal}, which can have a serious impact on the safety of both the mother and the fetus \cite{joo2021effect}. According to the World Health Organization (WHO)'s report on trends in maternal mortality, the number of maternal deaths worldwide in 2020 was approximately 28.7 thousand cases \cite{liu2024wearable}. In 2022, the global stillbirth rate was 13.9 per 1,000 births, meaning that for every 72 births, there was one stillbirth \cite{park2024traditional}. Fetal growth restriction (FGR) is a common pregnancy complication associated with various adverse pregnancy outcomes. It is a leading cause of stillbirth, neonatal death, and short- and long-term neonatal morbidity worldwide \cite{pels2020early} \cite{melamed2019neurodevelopmental} \cite{murray2015differential}. Despite improvements in neonatal care, FGR continues to increase the rate of neonatal mortality and short-term morbidity, and growth-restricted infants are at risk of long-term complications \cite{melamed2021figo} \cite{baschat2014neurodevelopment} \cite{figueras2011neurobehavioral}. Fetal growth depends not only on genetic factors, but also on various maternal, fetal and placental variables \cite{gantenbein2022highlighting}. Children of women with gestational diabetes mellitus (GDM) are often overweight \cite{voerman2019maternal} \cite{baidal2016risk}, particularly at the ages of 5 to 9 \cite{leth2025effect}, which increases the risk of lifestyle diseases \cite{ayer2015lifetime}. Therefore, the health of pregnant women affects fetal development, and fetal development determines the occurrence of adverse pregnancy outcomes. Early detection of fetal growth abnormalities can effectively reduce the risk of stillbirth \cite{moraitis2014birth} \cite{gardosi2013maternal} and ensure a positive pregnancy outcome \cite{boulvain2016induction}.
    
    Currently, the standard practice for identifying fetal growth abnormalities is to use the Hadlock formula to estimate fetal biometry based on ultrasound. However, due to differences in patient characteristics and ultrasound equipment quality, there is still considerable variation in the accuracy of scans \cite{andreasen2020multicenter}. Furthermore, this method is complex and costly \cite{khalil2017cerebroplacental} \cite{cavallaro2018using}, requiring specialised equipment and technicians, and is difficult to implement in resource-poor areas \cite{revathy2023pregnancy} \cite{liu2024wearable}. For example, in underdeveloped countries, due to the lack of integration in the healthcare system to facilitate information exchange, most pregnant women cannot undergo frequent prenatal check-ups during the early stages of pregnancy, increasing the risk of maternal and fetal mortality in these regions. Additionally, monitoring pregnant women's vital signs without medical supervision is impractical, and this situation is jeopardising the health of both mothers and fetuses \cite{monika2024analysis}. Furthermore, this traditional method cannot provide real-time fetal health monitoring and fails to meet pregnant women's demand for real-time fetal health monitoring.

    CTG is an external electronic monitoring of fetal heart rate (FHR) and uterine activity performed using an abdominal monitor, providing critical information about fetal accelerations, decelerations, baseline heart rate, and heart rate variability \cite{petrozziello2019multimodal} \cite{xie2024ai}. Since the 1970s, this technology has been clinically used to assess fetal health status \cite{david2022clinical} \cite{lin2024deep}. Additionally, CTG signals obtained through monitors offer the advantages of convenience, home monitoring, and low cost. However, most current research focuses on using CTG signals to monitor current fetal health status, with insufficient research on predicting future adverse outcomes. Existing studies have almost exclusively simplified CTG signals into a binary classification task for a specific adverse pregnancy outcome, neglecting the clinically more relevant multi-outcome scenarios \cite{ben2023computerized}, and lacking a generalized metric for predicting adverse pregnancy outcomes using CTG signals.

    Fetal chronological age refers to the time elapsed from the first day of the last menstrual period (LMP) to the present, and it serves as a crucial indicator for assessing fetal developmental stages. Meanwhile, biological age (of the fetus), typically estimated from fetal ultrasound examinations, reflects the actual developmental maturity of the fetus. The discrepancy between these two ages, which indicates whether fetal development is abnormal, may be significantly associated with adverse pregnancy outcomes.

\begin{figure}[t] 
        \centering 
        \includegraphics[width=\textwidth]{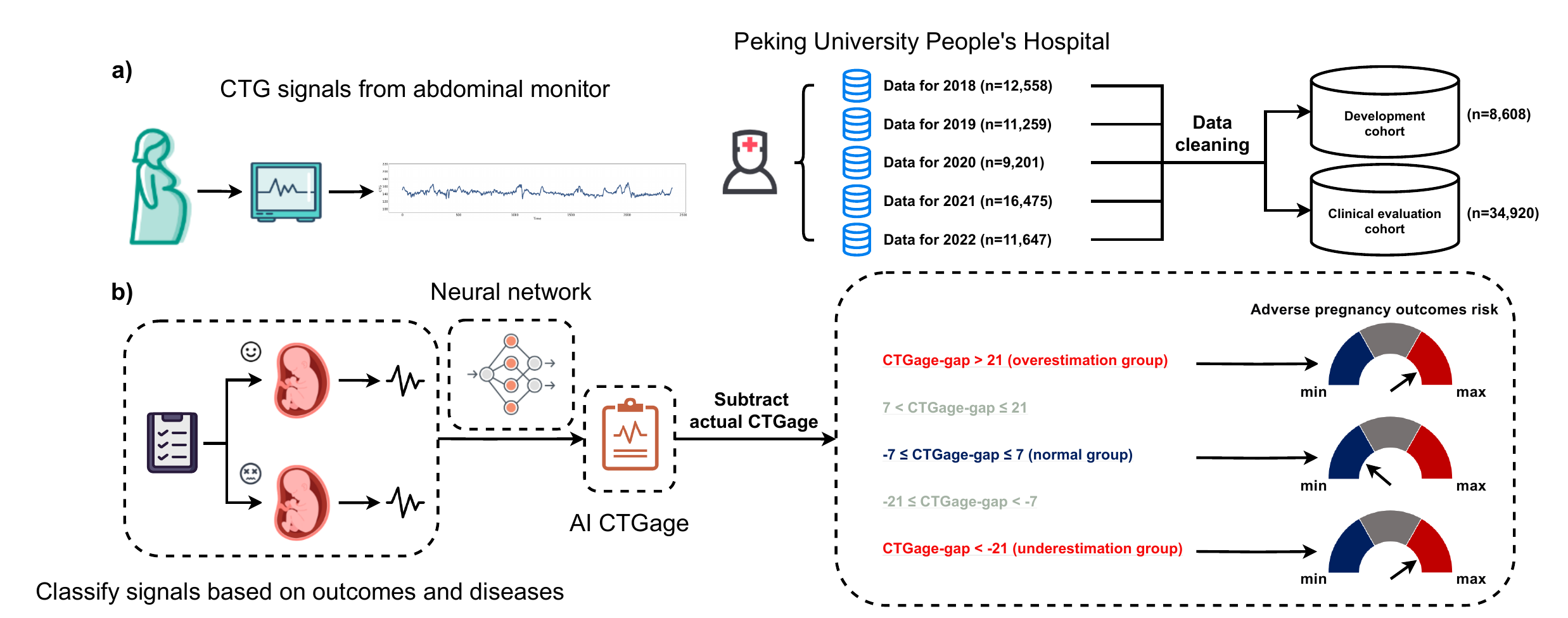} 
        \caption{Model development and application. a) We use CTG signals collected through abdominal monitor from Peking University People's Hospital and clean the data. b) We use a deep learning model to predict CTGage directly from CTG signals, then calculate the CTGage-gap and analyze its relationship with adverse pregnancy outcomes.} 
        \label{fig:pipeline} 
    \end{figure}

    In this pioneering work, we are the first to introduce CTGage—a fetal biological age predicted by an AI model from CTG time series data that functions as a novel, non-invasive, and easily accessible digital biomarker for indicating adverse pregnancy outcomes. We explore the potential association between CTGage-gap and adverse pregnancy outcomes by analyzing CTG time series, aiming to provide a new theoretical foundation and research perspective for CTG-based prediction of such outcomes. We also develop a deep learning method integrating data augmentation techniques to enhance data diversity (Figure \ref{fig:pipeline}), which not only positions CTGage as a novel digital biomarker for predicting adverse pregnancy outcomes but also addresses the current lack of a universal indicator in this field. Finally, clinical validation confirms CTGage’s efficacy as a digital biomarker, demonstrating its practical value in predicting adverse pregnancy outcomes using CTG time series.
    
\section*{RESULTS}


    \subsection*{Performance of AI CTGage Prediction Model}
\begin{figure}[h] 
            \centering 
            \includegraphics[width=0.8\textwidth]{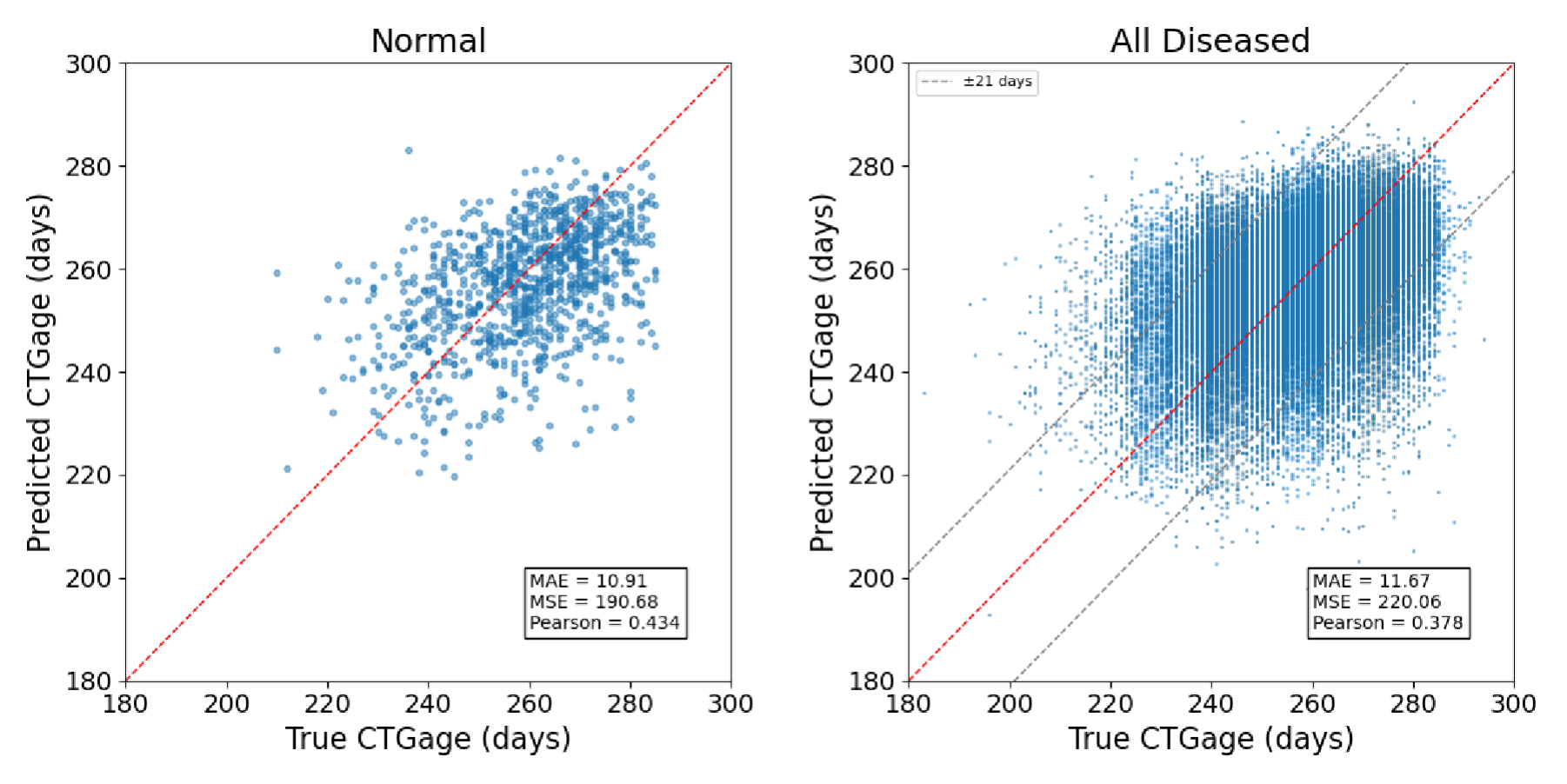} 
            \caption{Scatter plot comparison between normal and diseased samples (normal samples on the left, diseased samples on the right)} 
            \label{fig:散点图} 
        \end{figure}

        The evaluation indicators of the model are shown in Table \ref{tab:evaluation}, and comparison of normal samples and diseased samples on a scatter plot are shown in Figure \ref{fig:散点图}. From the scatter plot and evaluation indicators, we can see that there is a certain linear relationship between the actual CTGage and the AI CTGage, which indicates the effectiveness of using CTG signals in combination with the AI model to predict CTGage. At the same time, the comparison results between normal samples and diseased samples show that the prediction results of diseased samples are slightly worse than those of normal samples.

        \begin{table}[h]
            \centering
            \caption{Comparison of normal and diseased samples in terms of evaluation indicators}
            \begin{tabular}{cccccccc}
                \toprule
                 & Normal &  &  & Diseased &  &  \\
                \midrule
                 &  & 2018 & 2019 & 2020 & 2021 & 2022 & all diseased\\
                \midrule
                MAE  & \textbf{10.91} & 11.51 & 11.54 & 11.48 & 11.58 & 12.11 & 11.67\\
                MSE & \textbf{190.68} & 213.92 & 212.76 & 213.26 & 216.06 & 239.33 & 220.06\\
                Pearson & \textbf{0.434} & 0.356 & 0.371 & 0.382 & 0.408 & 0.378 & 0.378\\
                \bottomrule
            \end{tabular}
            \label{tab:evaluation}
        \end{table}

        AI CTGage can reflect the current development of the fetus, while CTGage-gap can reflect deviations in fetal development. Comparing the experimental results of the normal and diseased samples, we can see that in the normal samples, the absolute value of CTGage-gap is generally small, indicating that there are few cases of serious deviations in fetal development in the normal samples. However, in the diseased samples, there are cases where the absolute value of CTGage-gap is large, indicating that there are serious deviations in fetal development in these samples.
        
        If CTGage-gap is too large or too small, it indicates that the current development of the fetus is seriously inconsistent with the actual CTGage, which can lead to adverse pregnancy outcomes. At the same time, certain diseases in the mother can also affect the normal development of the fetus, resulting in abnormal CTGage-gap. Therefore, we will next analyze the relationship between CTGage-gap and adverse pregnancy outcomes, and between CTGage-gap and maternal diseases.

    \subsection*{Relationship between CTGage-gap and Adverse Pregnancy Outcomes}
        Table \ref{tab:zhanbi} shows that there are significant differences in the proportion of premature infants and low birth weight infants between different CTGage-gap groups. In terms of proportion, premature infants are mainly concentrated in the group with a CTGage-gap greater than 21 days, indicating that when fetal development is too abnormal, the risk of preterm birth is greater. Low birth weight infants, neonatal asphyxia, fetal distress, and fetal malformations are more likely to occur in the group with a CTGage-gap of less than -21 days, indicating that when fetal development is too abnormal, the risk of these adverse pregnancy outcomes is greater. As for neonatal congenital heart disease, it is not significantly correlated with CTGage-gap, either in terms of significance or proportion.

        \begin{table}[h]
            \centering
            \caption{Statistics on the proportion of adverse pregnancy outcomes in different CTGage-gap groups}
            \begin{tabular}{cccccc}
                \toprule
                \makecell{Outcomes} & \makecell{CTGage-gap\\ less than -21 \\(n=2930)}&\makecell{CTGage-gap\\ between -7 and 7\\(n=13152)}& \makecell{CTGage-gap\\ greater than 21 \\(n=2308)}& T test & H test\\
                \midrule\\
                \makecell{Premature \\infants, n (\%)} & 16 (0.55) & 187 (1.42) & 123 (5.33) & \textbf{<0.001} & \textbf{<0.001}\\ \\
                \makecell{Low birth weight\\ infants, n (\%)} & 5 (0.17) & 20 (0.15) & 3 (0.13) & 0.173 & \textbf{0.017}\\ \\
                \makecell{Neonatal \\asphyxia, n (\%)} & 33 (1.13) & 135 (1.03) & 23 (1.00) & 0.164 & 0.312\\ \\
                \makecell{Fetal distress,\\ n (\%)} & 20 (0.68) & 95 (0.72) & 9 (0.39) & 0.228 & 0.465\\ \\
                \makecell{Malformations,\\ n (\%)} & 23 (0.79) & 74 (0.56) & 15 (0.65) & 0.747 & 0.595\\ \\
                \makecell{Congenital heart\\ disease, n (\%)} & 18 (0.61) & 84 (0.64) & 16 (0.69) & 0.997 & 0.806\\ \\
                \bottomrule
            \end{tabular}
            \label{tab:zhanbi}
        \end{table}

\begin{figure}[h] 
            \centering 
            \includegraphics[width=\textwidth]{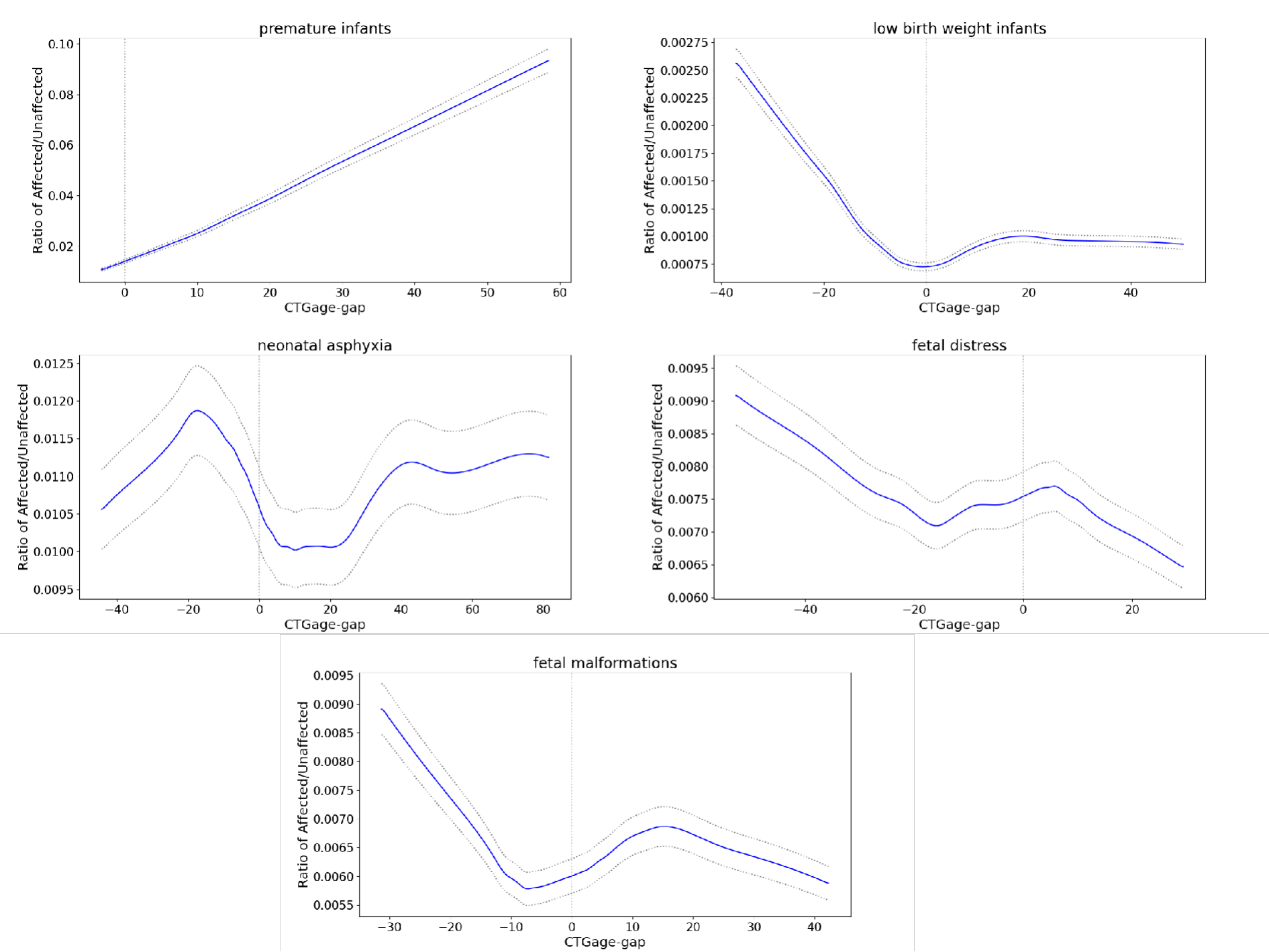} 
            \caption{Risk trends of adverse pregnancy outcomes in relation to CTGage-gap.} 
            \label{fig:risk} 
        \end{figure}

        As shown in Figure \ref{fig:risk}, the risk of premature infants increases with increasing CTGage-gap, while the risk of low birth weight infants, fetal distress, and fetal malformation increases with decreasing CTGage-gap. The risk of neonatal asphyxia increases with increasing absolute CTGage-gap. In summary, excessive or insufficient CTGage-gap may increase the risk of adverse pregnancy outcomes.
        
\begin{figure}[h] 
            \centering 
            \includegraphics[width=\textwidth]{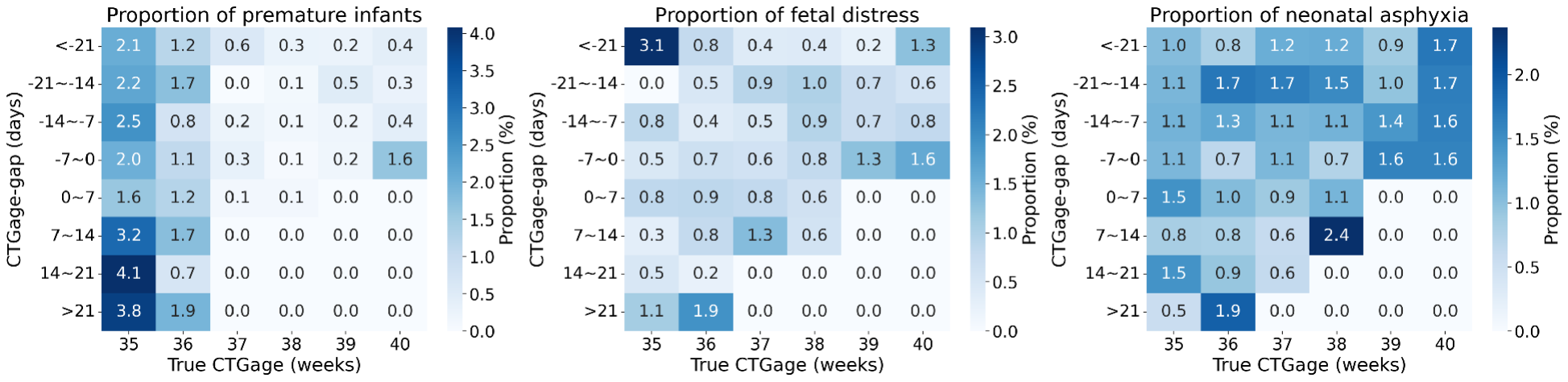} 
            \caption{Heat map of the distribution of adverse outcomes by CTGage-gap and actual CTGage.} 
            \label{fig:hotmap} 
        \end{figure}

        As shown in Figure \ref{fig:hotmap}, premature infants and fetal distress are concentrated in the actual CTGage range of 35 to 36 weeks, indicating that the risk of these two adverse pregnancy outcomes is higher at this stage. Neonatal asphyxia accounts for a significant proportion of cases in the actual CTGage range of 35 to 40 weeks, indicating that the risk exists throughout the entire stage.

    \subsection*{Relationship between CTGage-gap and Maternal Diseases }
        The results of the study indicate that maternal diseases may affect the normal development of the fetus, leading to CTGage-gap. Specifically, certain maternal diseases such as gestational diabetes mellitus (GDM), anaemia, maternal congenital disease, umbilical cord problems, and placental lesions can interfere with the normal growth and development of the fetus. As shown in Table \ref{tab:zhanbi_m}, GDM may cause excessive fetal growth, resulting in AI CTGage being higher than actual CTGage and producing a positive CTGage-gap. While diseases such as anaemia may cause restricted fetal growth, resulting in AI CTGage being lower than actual CTGage and producing a negative CTGage-gap. Maternal congenital disease may cause excessive or insufficient CTGage-gap and needs to be analyzed in combination with specific congenital diseases.

        \begin{table}[h]
            \centering
            \caption{Statistics on the proportion of maternal diseases in different CTGage-gap groups}
            \begin{tabular}{cccccc}
                \toprule
                \makecell{Outcomes} & \makecell{CTGage-gap\\ less than -21 \\(n=2930)}&\makecell{CTGage-gap\\ between -7 and 7\\(n=13152)}& \makecell{CTGage-gap\\greater than 21 \\(n=2308)}& T test & H test\\
                \midrule\\
                \makecell{GDM, n (\%)} & 450 (15.36) & 2744 (20.86) & 737 (31.93) & \textbf{<0.001} & \textbf{<0.001}\\ \\
                \makecell{Anaemia, n (\%)} & 1099 (37.51) & 4569 (34.74) & 703 (30.46) & \textbf{<0.001} & \textbf{<0.001}\\ \\
                \makecell{Maternal\\congenital \\disease, n (\%)} & 8 (0.27) & 23 (0.17) & 6 (0.26) & 0.460 & 0.340\\ \\
                \makecell{Umbilical cord \\problems, n (\%)} & 1162 (39.66) & 5125 (38.97) & 864 (37.44) & \textbf{0.023} & 0.282\\ \\
                \makecell{Placental\\ lesions, n (\%)} & 441 (15.05) & 1905 (14.48) & 337 (14.60) & 0.356 & 0.163\\ \\
                \bottomrule
            \end{tabular}
            \label{tab:zhanbi_m}
        \end{table}


        In summary, maternal diseases affect the growth rate of the fetus, leading to a difference between the AI CTGage and the actual CTGage, which reflects a mismatch between fetal development and actual CTGage, thereby increasing the risk of adverse pregnancy outcomes.

    \subsection*{Interpretable Visualization Analysis }
            We visualize the model's attention to CTG signals for different CTGage-gap groups. We calculate the gradients of the input data relative to the model output, which reflect the contribution of each point in the input data to the model output. We then apply a logarithmic transformation to enhance the contrast of the gradients and use Gaussian filtering to smooth the attention weights. Finally, we normalize the gradients to the range (0,1) for use in color mapping. During prediction, the red areas receive more attention from the model.

\begin{figure}[htbp]
                \centering
                \includegraphics[width=\linewidth]{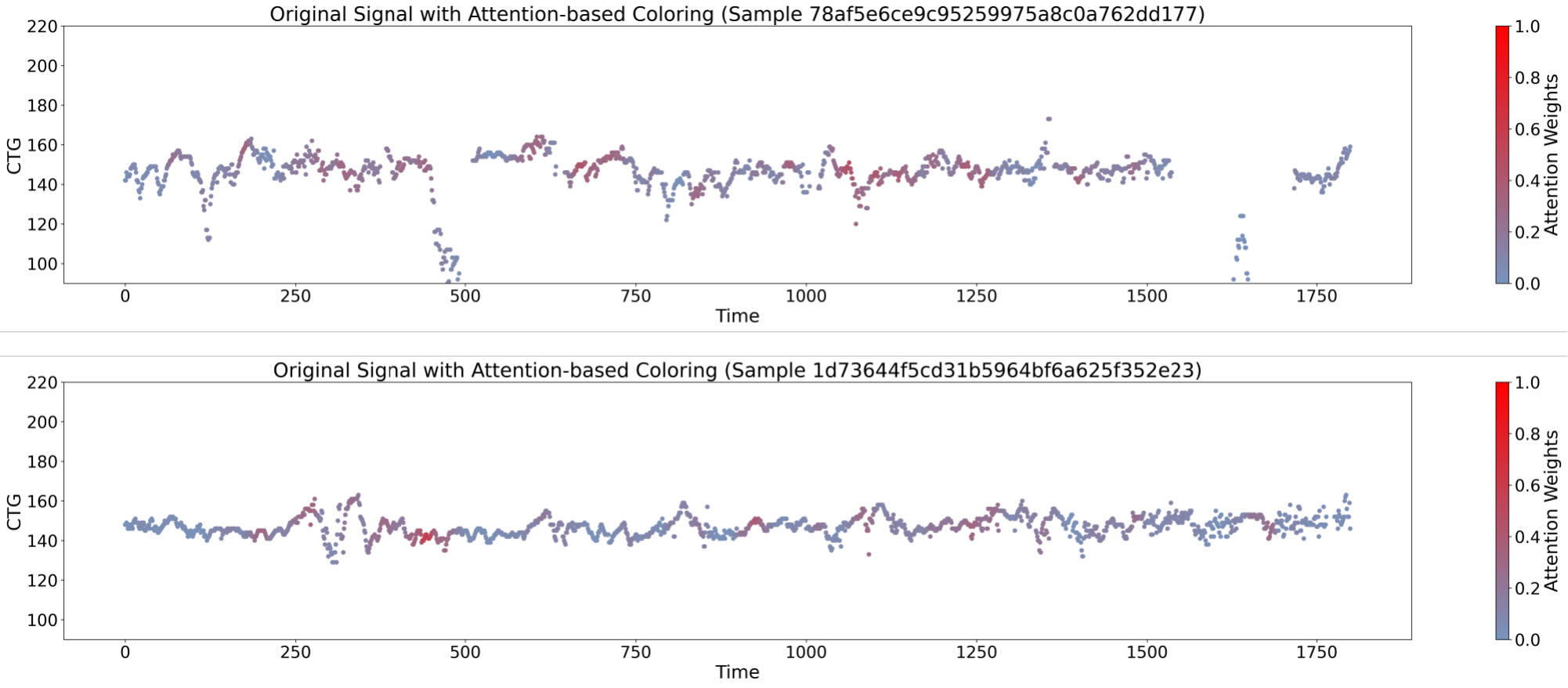}
                \caption{CTG attention visualization of neonatal asphyxia samples.}
                \label{fig:zhixi}
            \end{figure}

\begin{figure}[htbp]
                \centering
                \includegraphics[width=\linewidth]{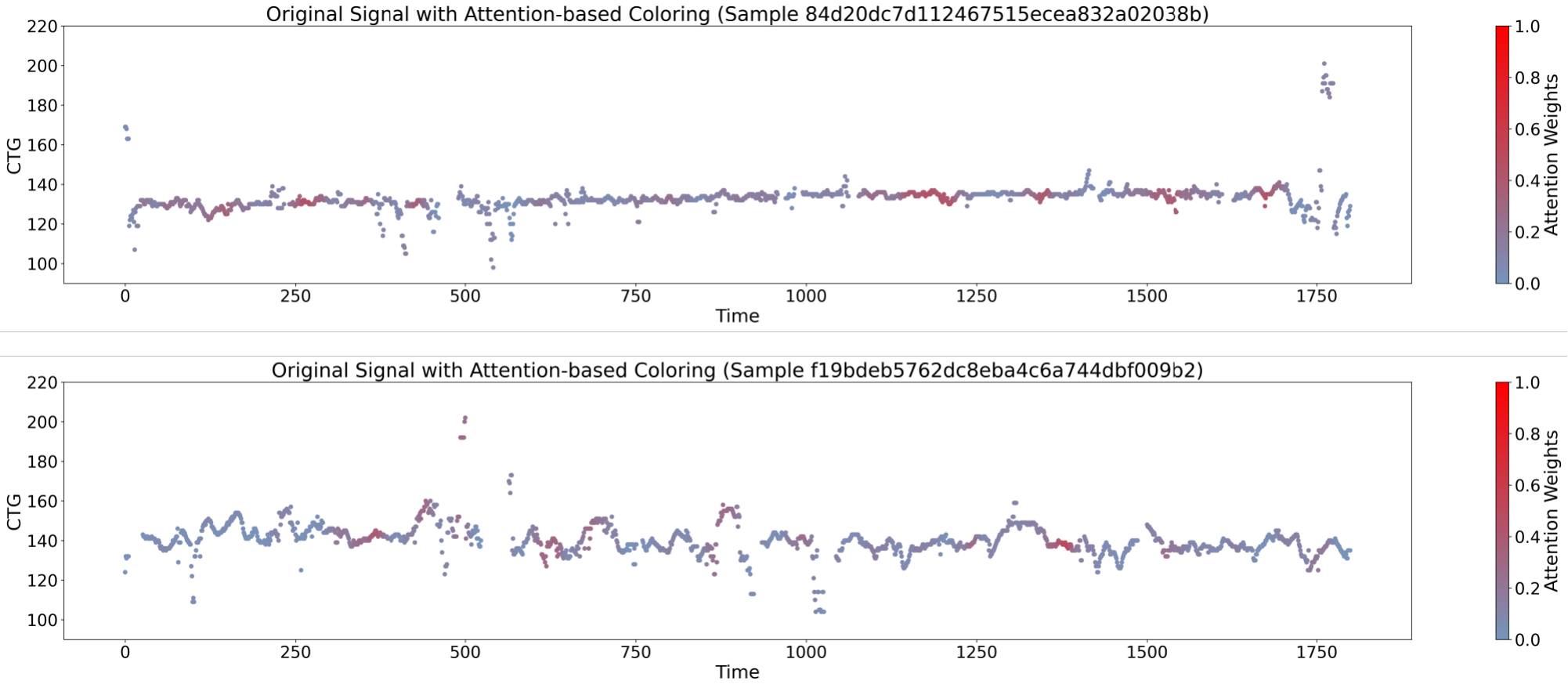}
                \caption{CTG attention visualization of premature infants samples.}
                \label{fig:zaochan}
            \end{figure}

\begin{figure}[htbp] 
                \centering 
                \begin{subfigure}{\linewidth}
                    \centering
                    \includegraphics[width=\linewidth]{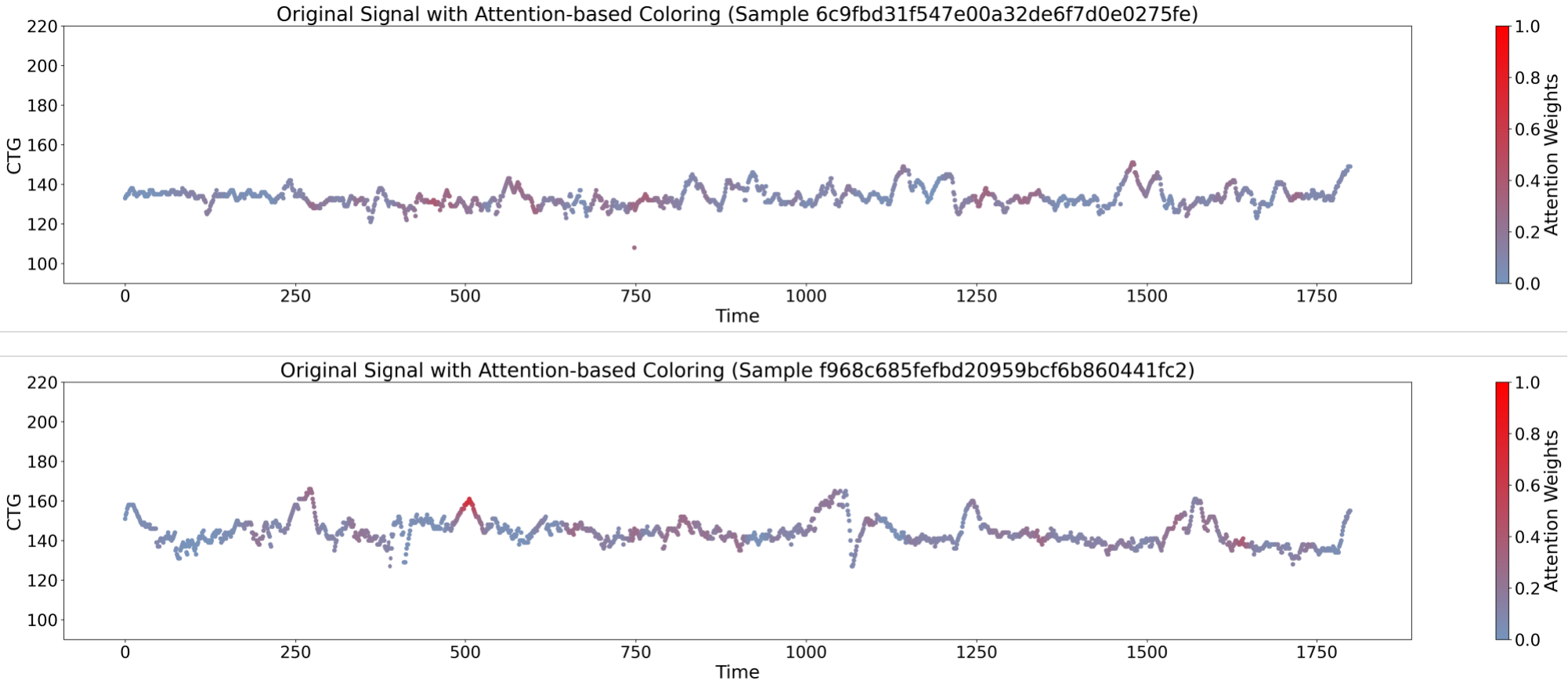}
                    \caption{True negative samples}
                    \label{fig:normal-true}
                \end{subfigure}
                
                
                \begin{subfigure}{\linewidth}
                    \centering
                    \includegraphics[width=\linewidth]{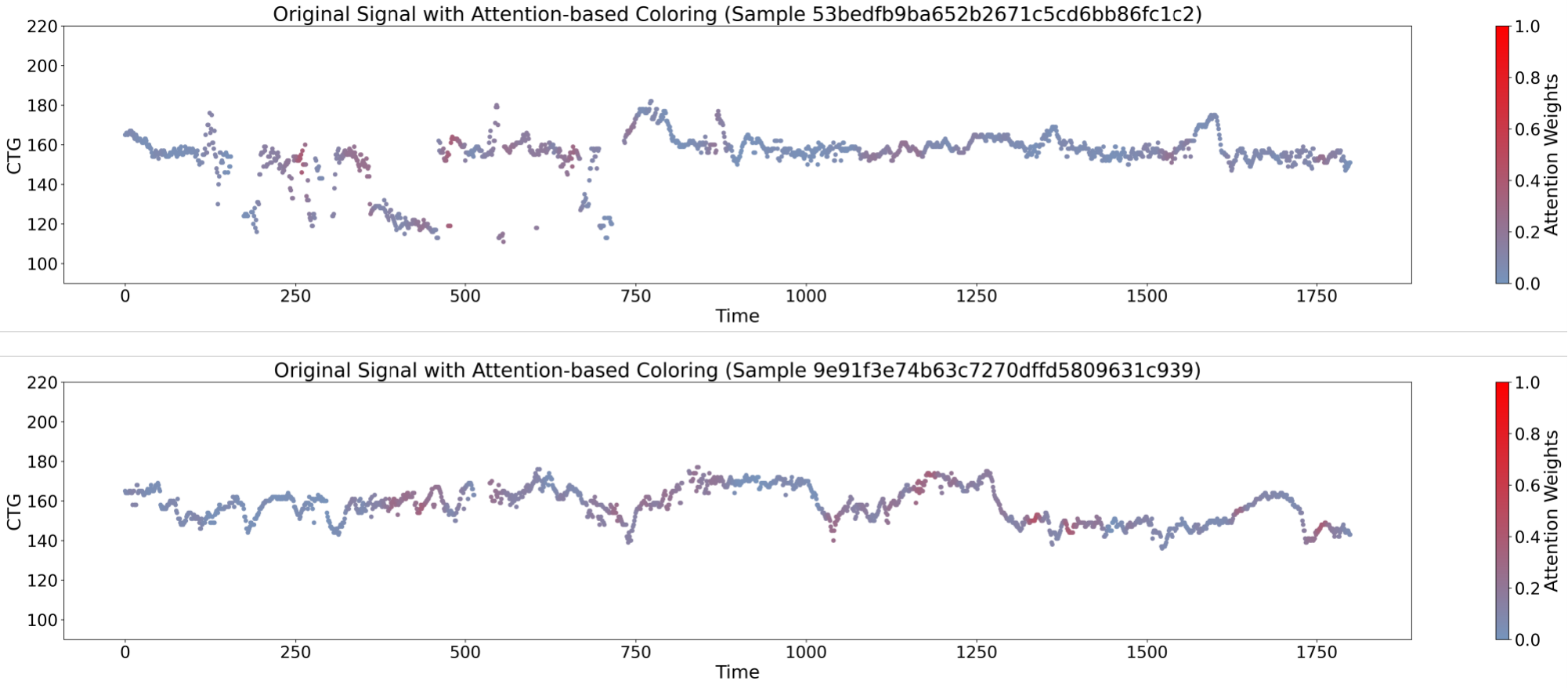}
                    \caption{False positive samples}
                    \label{fig:normal-false}
                \end{subfigure}
                \caption{CTG attention visualization of samples without adverse pregnancy outcomes.} 
                \label{fig:attention_non} 
            \end{figure}

            Figure \ref{fig:zhixi} shows two pregnant women with neonatal asphyxia. The model outputs CTGage-gap of -34.30 and -44.75 for these two samples, both of which belong to the underestimation group. Figure \ref{fig:zaochan} shows two pregnant women with premature infants. The model outputs CTGage-gap of 31.58 and 38.27 for these two samples, both of which belong to the overestimation group. Figure \ref{fig:attention_non} shows four pregnant women who did not have adverse pregnancy outcomes. The CTGage-gap outputs in Figure \ref{fig:normal-true} are -4.39 and 1.92, respectively, which belong to the normal group. But the CTGage-gap outputs in Figure \ref{fig:normal-false} are 38.41 and -41.67, respectively, which belong to developmental abnormalities and are false positives.

            \begin{figure}[htbp] 
                \centering 
                \begin{subfigure}{\linewidth}
                    \centering
                    \includegraphics[width=\linewidth]{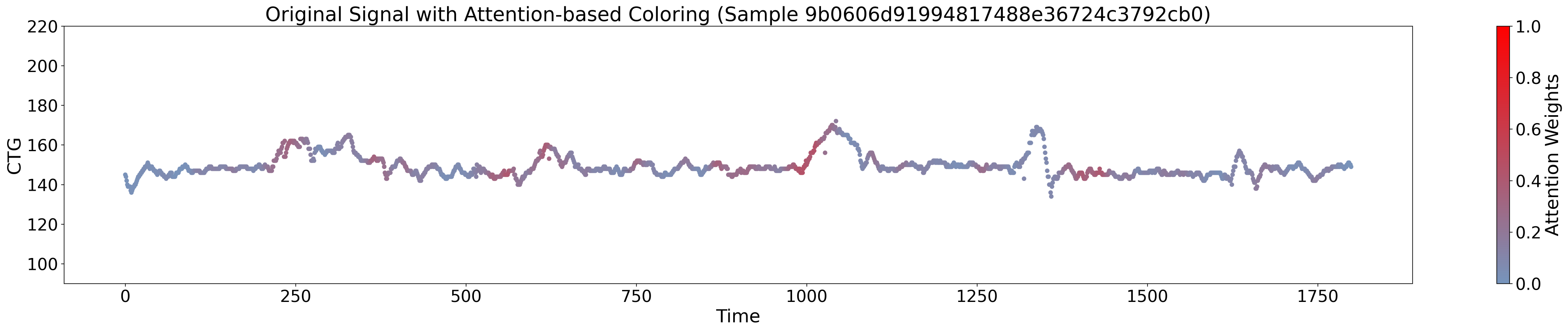}
                    \caption{The AI CTGage is 248.78, the actual CTGage is 251, and the CTGage-gap is -2.22, indicating that the fetus is developing normally at this time.}
                    \label{fig:case251}
                \end{subfigure}

                \begin{subfigure}{\linewidth}
                    \centering
                    \includegraphics[width=\linewidth]{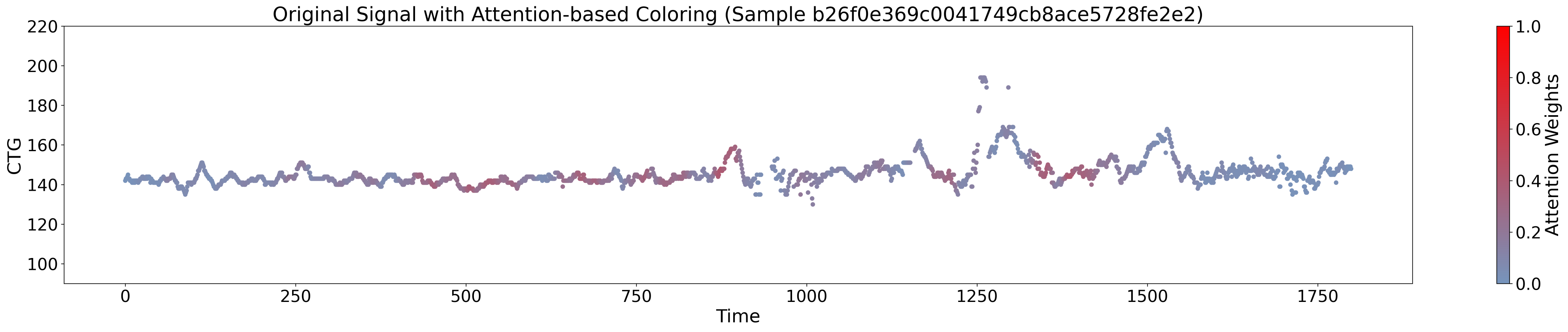}
                    \caption{The AI CTGage is 242.31, the actual CTGage is 268, and the CTGage-gap is -25.69. At this time, the fetus is already in a high-risk situation.}
                    \label{fig:case268}
                \end{subfigure}

                \begin{subfigure}{\linewidth}
                    \centering
                    \includegraphics[width=\linewidth]{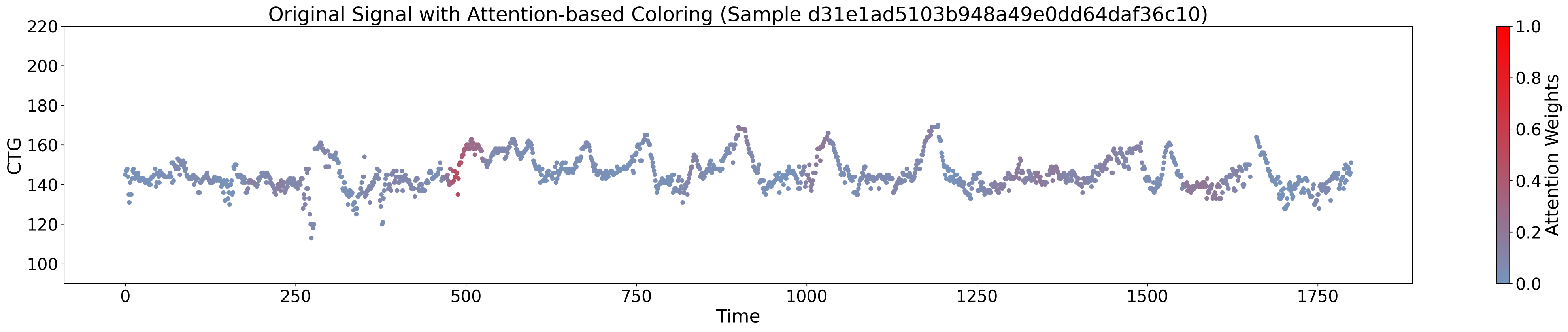}
                    \caption{The AI CTGage is 248.81, the actual CTGage is 278, and the CTGage gap is -29.19. At this time, the fetus is still in a high-risk situation.}
                    \label{fig:case278}
                \end{subfigure}

                \begin{subfigure}{\linewidth}
                    \centering
                    \includegraphics[width=\linewidth]{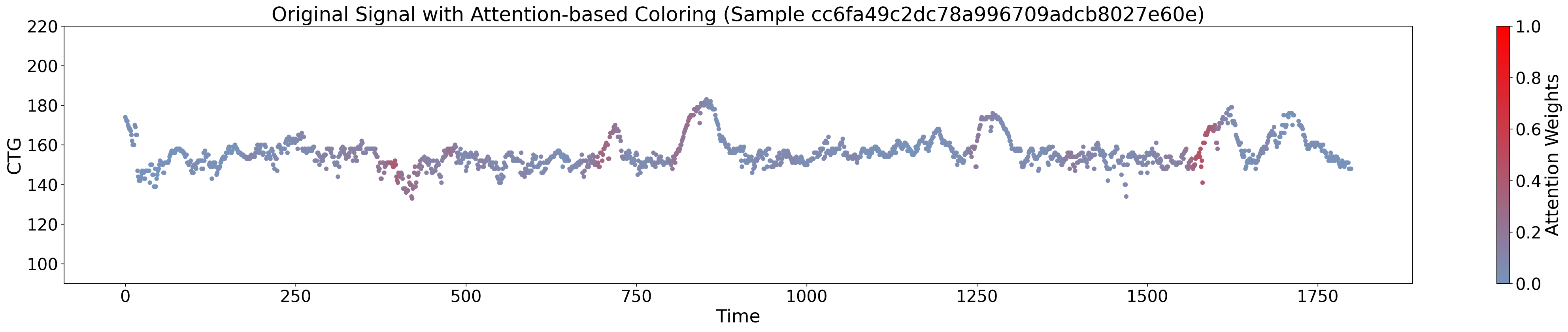}
                    \caption{The AI CTGage is 250.02, the actual CTGage is 282, and the CTGage-gap is -31.98. At this time, the fetus is still in a high-risk situation, and the CTGage-gap is becoming increasingly severe.}
                    \label{fig:case282}
                \end{subfigure}
                \caption{CTG attention visualization of a pregnant woman under different actual CTGage. The pregnant woman suffered from hypothyroidism during pregnancy, and finally had adverse pregnancy outcomes of fetal malformation.} 
                \label{fig:case} 
            \end{figure}

            Figure \ref{fig:case} shows the visualization of CTG attention for a pregnant woman (No. 5073232) under different actual CTGages. It can be seen that when the actual CTGage is 251 days, the fetus is developing normally, and the overall trend of CTG signal is stable. However, starting from the second recording (actual CTGage at 268 days), fetal development began to show abnormalities, and CTG signals began to experience deceleration events accompanied by an increase in attention weights. When the actual CTGage is 278 days, fetal development still shows abnormalities, with an increased range of CTG signal fluctuations and high variability. When the actual CTGage is 282 days, the CTG signal fluctuates greatly, showing extremely high variability. From the results of CTGage-gap, it can be seen that the fetal development continued to deteriorate and became increasingly severe in the second to fourth recordings.
            
            The pregnant woman suffers from hypothyroidism, which can lead to fetal developmental abnormalities and increase the risk of adverse pregnancy outcomes if not treated or managed in a timely manner. Ultimately, the pregnant woman experienced an adverse pregnancy outcome with fetal malformations and chose to terminate the pregnancy.

            Through this case, we can find that the model highlights key areas in CTG signals through attention mechanisms, such as deceleration, acceleration, and regions with high variability, which may be closely related to fetal development. This attention-based analysis can help clinical doctors quickly identify abnormal parts in CTG signals, enabling earlier intervention and evaluation.

\section*{DISCUSSION}
    This study proposes and validates a new AI-based CTG analysis method, using AI-derived CTGage as a digital biomarker to predict adverse pregnancy outcomes. The results show that the larger the absolute value of CTGage-gap, the higher the risk of adverse pregnancy outcomes for the fetus. This suggests that AI-derived CTGage could serve as a potential indicator for early identification of high-risk pregnancies.

    This study found that fetuses with CTGage-gap exceeding ±21 days had a significantly higher incidence of adverse outcomes. This is consistent with previous studies, which found that fetal growth deviation from the actual CTGage (such as FGR or excessive growth) is itself an important predictor of adverse pregnancy outcomes \cite{crispi2018long}. In addition, maternal diseases such as GDM, anaemia, and placental lesions can also cause fetal growth to deviate from the normal trajectory by affecting placental perfusion or the metabolic environment, resulting in abnormal CTGage-gap \cite{melamed2021figo}. This study uses growth curves to confirm that there is a certain pattern in the changes in adverse pregnancy outcomes in CTGage-gap, namely, the greater the absolute value of CTGage-gap, the greater the risk of disease. And this study also confirms significant differences in CTGage-gap distribution under different maternal disease conditions through box plots. Maternal disease may cause CTGage-gap abnormalities by affecting the fetal growth environment, indicating that CTGage-gap abnormalities can be a potential signal of the impact of maternal disease on the fetus. At the same time, this study further reveals significant differences between different CTGage-gap groups in the CTG feature space through t-SNE visualization analysis, indicating that the AI-derived CTGage outputted by the model can be used as an effective indicator to distinguish whether fetal development deviates from the actual CTGage.

    This study adopts a Net1D architecture combined with distribution-aligned augmented regression to effectively mitigate the bias caused by sample distribution imbalance in CTGage prediction. Compared to traditional regression models, this method demonstrates stronger robustness in tail samples, significantly improving prediction accuracy. Additionally, the sliding window and random perturbation data augmentation strategies significantly improve the model's generalization ability across different CTGage segments. More importantly, we use gradient visualization methods to identify the critical time periods the model focuses on in CTG sequences, providing new insights for enhancing model interpretability and offering potential solutions to the 'black box problem' in future clinical deployments.

    Traditional fetal health assessments rely on ultrasound and biochemical indicators, which are highly accurate but costly and difficult to access, especially in resource-limited areas. In contrast, CTG devices are portable, low-cost, and can be used at home. Combined with AI models, they can achieve continuous, non-invasive, and scalable fetal health monitoring. This study marks the first time CTG data has been expanded from ‘real-time status assessment’ to ‘long-term risk prediction,’ breaking through its traditional application scenarios and holding significant public health implications.

    Although this study has a large sample size and good model performance, there are still limitations. Firstly, the data source is single, and there may be population bias issues, which will require multi-central and multi-ethnic verification in the future. Secondly, the definition of adverse outcomes is limited. This study only included six types of adverse fetal outcomes and several maternal complications, and did not cover more serious outcomes. Thirdly, although the CTGage-gap threshold of ±21 days is set based on clinical experience, further prospective studies are needed to validate the optimal cutoff point. Finally, while the model's interpretability has been preliminarily explored, it has not yet achieved rule outputs that are directly understandable by clinicians, necessitating the introduction of more interpretable AI methods in the future.

    
    Future research will focus on the following directions. Firstly, validate the model's generalization ability in larger-scale, multi-ethnic, multi-device source data. Secondly, dynamically adjust the risk stratification criteria for CTGage-gap based on prospective cohort studies. Thirdly, develop a lightweight model that can be embedded in CTG devices to enable remote monitoring and risk warning during pregnancy. Finally, combine multi-modal data such as placental ultrasound and maternal metabolomics to deeply analyze the biological basis of CTGage-gap abnormalities.
    

\section*{METHODS}

    \subsection*{Dataset and Preprocessing}
        \textbf{Dataset} \: 
        The dataset of this study covers detailed obstetric information collected from Peking University People's Hospital between 2018 and 2022, aiming to provide comprehensive data support for the assessment of fetal health and the study of adverse pregnancy outcomes. The dataset contains 61,140 CTG signals from 11,385 subjects, and data are collected at a frequency of 2 Hz, with each data point representing a 1-minute fetal heartbeat value. The dataset is collected without any pre-processing steps, preserving the integrity of the original data. Each record also contains detailed information about the pregnant woman, pregnancy monitoring data, labor and delivery, and newborn health indicators. This study was approved by the Institutional Review Board of Peking University People's Hospital (approval number 2022PHB169-001). This study adhered to the Declaration of Helsinki.

        The basic information in the dataset includes maternal age, gestational age, monitoring time, medical record number, and discharge diagnosis, among others. The maternal age range is 15 to 54 years old, with an average age of approximately 33 years. This information provides the foundation for patient identification and timeline references for the study.  In addition, the dataset also records information such as fetal malformations, fetal distress, neonatal asphyxia, and low birth weight infants, which can help evaluate the health status of newborns and the medical interventions they may need. Finally, the dataset also covers the mother's health status, including gestational diabetes mellitus (GDM), maternal congenital diseases, and thyroid function indicators. These data help assess the impact of maternal health status on pregnancy outcomes.

        \textbf{Outcomes and disease definitions} \:     
        We extract multiple definitions of adverse pregnancy outcomes related to the fetus and maternal diseases from the dataset. Among these, adverse pregnancy outcomes related to the fetus include premature infants, low birth weight infants, neonatal asphyxia, fetal distress, fetal malformations, and neonatal congenital heart disease. Maternal-related conditions include gestational diabetes mellitus, anaemia, maternal congenital disease, placental lesions, and umbilical cord-related problems such as umbilical cord prolapse and umbilical cord entanglement.
        
\begin{figure}[t] 
            \centering 
            \includegraphics[width=0.8\textwidth]{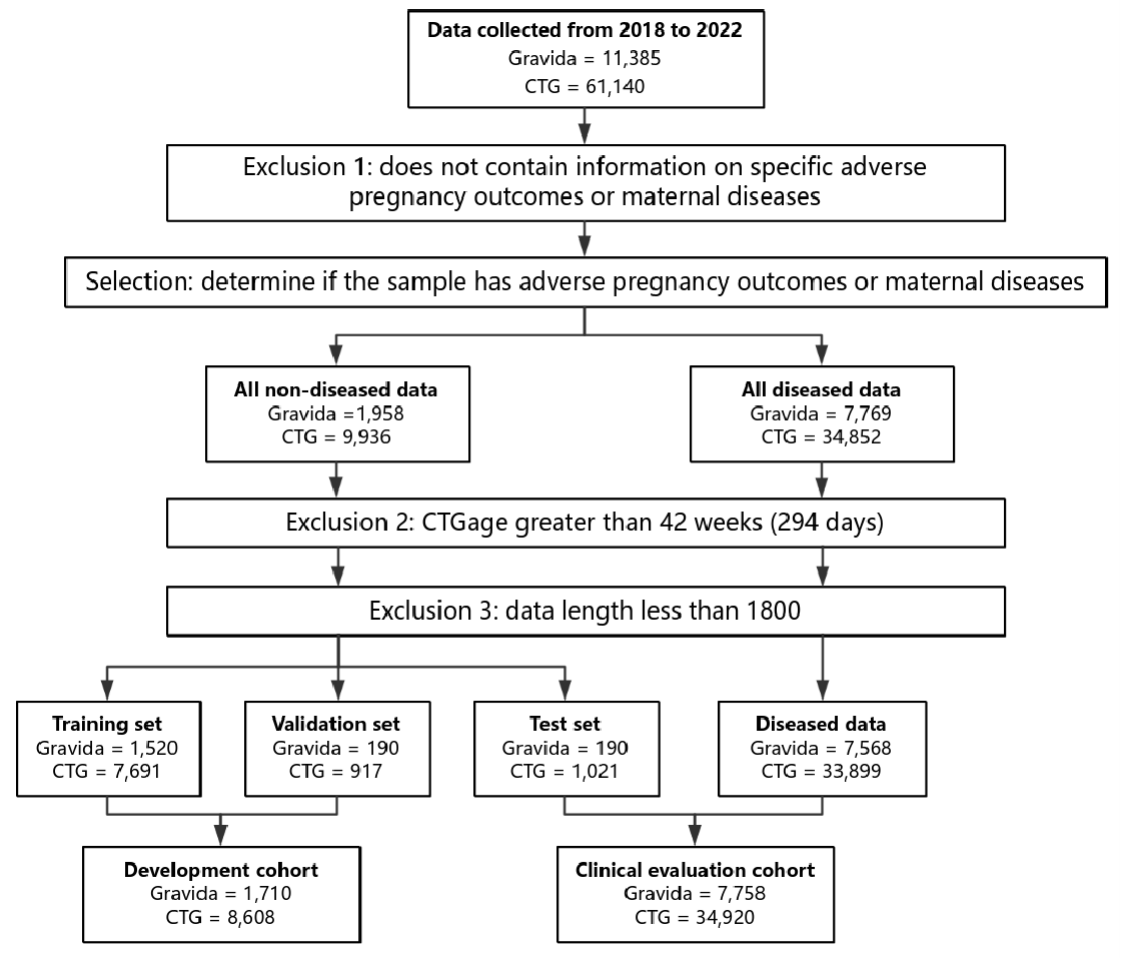} 
            \caption{Data flow chart. Data is screened according to three criteria and classified according to the presence or absence of adverse pregnancy outcomes or maternal diseases. Non-disease data is divided into training, validation, and test sets. The test set of non-disease data and disease data is used as the clinical evaluation cohort, while the training and validation sets of non-disease data are used as the development cohort.} 
            \label{fig:data_info} 
        \end{figure}

        \textbf{Preprocessing} \: 
        The division of the dataset is based on the presence or absence of adverse pregnancy outcomes or maternal diseases in pregnant women. The data flow is depicted in Figure \ref{fig:data_info}. Specifically, we divide the dataset into the development cohort and the clinical evaluation cohort. The development cohort contains a part of the records without adverse pregnancy outcomes and maternal diseases, while the clinical evaluation cohort contains the remaining normal records and records with adverse pregnancy outcomes or maternal diseases. During the segmentation process, we select several specific adverse pregnancy outcomes and maternal diseases and then segment the dataset based on the presence or absence of these. The normal dataset is divided into training, validation, and test sets in a ratio of 8:1:1. This segmentation ensures that the model can more accurately predict the CTGage of pregnant women and thus better observe the relationship between CTGage-gap and adverse pregnancy outcomes.

        We screen CTG signals with a data length of no less than 1,800 and the gestational age range is not less than 42 weeks. For adverse pregnancy outcomes, we extract premature infants, low birth weight infants, neonatal asphyxia, fetal distress, fetal malformations, and neonatal congenital heart disease. While for maternal diseases, we extract gestational diabetes mellitus, anemia, maternal congenital disease, placental lesions as well as adverse outcomes related to the umbilical cord including umbilical cord around the neck, umbilical cord around the body, and so on.

    \subsection*{Model Development}
        \subsubsection*{Model Architecture}
            The CTGage prediction model is based on Net1D \cite{hong2020holmes}, a one-dimensional convolutional neural network (CNN) constructed on a variant of ResNet \cite{he2016deep}. It is chosen because of its ability to efficiently extract complex features from CTG signals and map them to the CTGage prediction task. Net1D employs sequential convolutional blocks with residual connectivity \cite{he2016deep}, which allows signals in the network to be passed directly bypassing some of the layers, helping to alleviate the problem of gradient vanishing in deeper networks and improving the stability of training. In addition, Net1D introduces Squeeze-and-Excitation Modules \cite{hu2018squeeze}, which enhance the network's focus on important features by adaptively recalibrating the channel weights, and can dynamically adjust the importance of the feature channels, thus improving the feature extraction capability of the model. This architecture ensures robust and efficient processing of CTG signals for CTGage prediction.
            
        \subsubsection*{Distribution-Aligned Augmented Regression in CTGage Prediction}
            To address the issue of data imbalance, we propose a distribution-aligned augmented regression technique. The core idea of this technique is to address prediction bias caused by the sparse distribution of samples under certain labels and sample distribution shifts by enhancing samples with sparse label distributions and aligning the predicted label distribution with the theoretical label distribution. This approach also further enhances data diversity.

            We first perform data augmentation on non-diseased samples using sliding window technology, and then further augment samples with sparse label distributions. Each original data point is subjected to random temporal distortion (stretching or compression) and random noise addition, followed by sliding window augmentation.

            Enhancing samples with sparse label distributions can only moderately alleviate the data imbalance issue, and in reality, prediction values still tend to cluster within the range of labels with larger sample sizes. To address this, we adopt the DistLoss \cite{nie2024deep} and made improvements. The core of this loss function is to enable the network not only to perform point-to-point label regression but also to align the distribution shape of batch predictions with the prior theoretical label distribution.

            To provide a prior distribution for DistLoss, we construct a truncated normal distribution using the gestational age of all samples in the training set. Specifically, we first calculate the mean $\mu$ and standard deviation $\sigma$ of the age in the training set, then multiply $\sigma$ by 0.8 to appropriately narrow the bandwidth. Then discretise the interval $[label_{min}, label_{max}]$ at intervals of $step$, and calculate the probability density function values. To avoid zero probability in sparse intervals, add a minimal constant $\epsilon$ to each discrete bin and renormalise to ensure the distribution sums to 1. The resulting vector is the theoretical label. During training, first perform a differentiable sorting on the network output to obtain a smooth cumulative distribution curve. Then use the L1Loss distance metric to measure the difference between the predicted label distribution and the theoretical label distribution, with the loss output being $L_{Dist}$. 
            \begin{equation}
                L_{Dist}=\frac{1}{N}\sum_{i=1}^{N}\vert \hat{y_i}^{\uparrow}-p_i\vert,
            \end{equation}
            where $\hat{y}^\uparrow$ denotes the ordered vector obtained by sorting the predicted values in a differentiable manner, and $p$ denotes the theoretical distribution vector. 
            Then, for each sample, the L1 Loss is calculated between the predicted value and the true value, ensuring that the absolute error for each sample is not sacrificed. The output loss is $L_{Point}$.
            \begin{equation}
                L_{Point}=\frac{1}{N}\sum_{i=1}^{N}\vert \hat{y_i}-y_i\vert,
            \end{equation}
            where $\hat{y}$ is the model prediction and $y$ is the true label.
            DistLoss requires the predicted label distribution to converge towards the theoretical label distribution as a whole, but if the tail weights of the theoretical label distribution are insufficient, the model will reduce the predicted values of a few tail samples in order to minimise the loss, resulting in higher labels being predicted lower. To address this, we set a monotonic regularisation to impose an additional penalty on the deviation of the regression slope between the predicted value and the actual value from 1, preventing the distortion trend where the predicted value decreases as the label increases. First, calculate the linear regression slope, then calculate the regularisation loss $L_{Slope}$.
            \begin{equation}
                \hat{\beta}=\frac{Cov(\hat{y},y)}{Var(\hat{y})+\epsilon},
            \end{equation}
            \begin{equation}
                L_{Slope}=(\hat{\beta}-1)^2,
            \end{equation}
            where $\epsilon$ is a small constant added for numerical stability.
            Finally, the three losses are weighted and summed to optimize the regression performance of the model.
            \begin{equation}
            L_{Total}=\lambda_{Dist}L_{Dist}+\lambda_{Point}L_{Point}+\lambda_{Slope}L_{Slope}
            \end{equation}
        \subsubsection*{Synergistic Effect of Distribution-Aligned Augmented Regression and Net1D Model}
            In this study, the synergistic effect of distribution-aligned augmented regression technology and the Net1D model significantly improved model performance. With the help of the Net1D model, we are able to calculate CTGage and derive key indicators of CTGage-gap. This indicator, as a digital biomarker, constitutes the core output of this study and provides an important basis for subsequent analysis of its association with adverse pregnancy outcomes.
        \subsubsection*{Training Details}
            The model is implemented using the PyTorch 2.0 framework and trained on a high-performance server equipped with an Intel Core i9-10980XE processor (3.00GHz) and four NVIDIA GeForce RTX 3090 graphics cards (each with 24GB of memory). Early stopping is used during training, which took a total of 192 epochs. The Adam optimiser is used with an initial learning rate of 1e-3. Additionally, L2 weight regularisation is applied with a regularisation factor $\lambda = 1e-3$ to reduce overfitting. Furthermore, Cosine Annealing with Warm Restarts is employed to help the model repeatedly escape local minima and converge to a more optimal solution.
    \subsection*{Model Interpretation}
        To enhance the interpretability of the model, we use the input gradient to generate attention weights and visualise the CTG signals with colour coding to reveal the key periods that the model focuses on in predicting CTGage. Specifically, the gradients of the input sequence on the model output are first calculated, and their absolute values are taken. These values are then subjected to logarithmic compression and normalisation to obtain the attention weights. Gaussian smoothing is then applied to remove noise, and the results are mapped to a colour bar, with darker colours indicating a greater influence of that segment on the prediction outcome. This approach allows for an intuitive observation of which intervals the model considers to be the critical regions for CTGage estimation.

    \subsection*{Statistical Analysis Methods for Clinical Validation}
        In the clinical validation of this study, we use a series of statistical analysis methods to evaluate the relationship between CTGage-gap and adverse pregnancy outcomes. We also analyze the impact of maternal disease on CTGage-gap.
        
        AI CTGage reflects the current development of the fetus, while CTGage-gap reflects whether the current fetal development is consistent with the actual CTGage. We divide the CTGage-gap into five groups according to ±7 and ±21 days: less than -21 days, -21 to -7 days, -7 to 7 days, 7 to 21 days, greater than 21 days. This is because a CTGage-gap within ±7 days is a normal physiological fluctuation range of developmental deviation that does not require special intervention, but if the fetal developmental deviation is greater than three weeks (21 days), this is usually beyond the normal range of variation in clinical practice, and there may be pathological developmental abnormalities that require timely investigation of potential pathological factors. Therefore, the first and fifth groups are more likely to be directly associated with adverse outcomes and are considered high-risk groups, with stronger analytical necessity and clinical significance. The third group represents normal development and serves as the normal reference group. The second and fourth groups fall into the ‘grey zone’ of developmental deviation, which can typically be managed through enhanced monitoring with lower short-term risks, and thus are not the primary focus of analysis. Therefore, we will focus on analyzing the experimental results of the first, third, and fifth groups.
        
        We analyze the incidence of adverse pregnancy outcomes in each group to understand the impact of different CTGage-gap on adverse pregnancy outcomes. To further verify the statistical significance of these differences, we use the T-test to compare the impact of CTGage-gap on adverse pregnancy outcomes and use the Kruskal-Wallis H test to compare the differences in adverse pregnancy outcomes between different CTGage-gap groups. In addition, we plot a curve to quantify the relationship between CTGage-gap and adverse pregnancy outcomes. At the same time, we also plot a box plot to compare the effect of maternal disease on CTGage-gap. All statistical test results are expressed as P values, where P values less than 0.05 are considered statistically significant. Together, these analyses provide comprehensive statistical support for evaluating the effectiveness of CTGage-gap as a potential numerical biomarker.

\section*{DATA AVAILABILITY}

The codes are open source and available at \url{https://github.com/PKUDigitalHealth/CTGage} or \url{https://github.com/Gss0217/CTGage}.
The data are available from the authors upon reasonable request.

\section*{ACKNOWLEDGMENTS}

    This work was supported by the National Natural Science Foundation of China (62102008), and National Natural Science Foundation of China Regional Innovation Development Joint Fund (U20A20388). 

\section*{DECLARATION OF INTERESTS}

    The authors declare no competing interests related to this work.

\newpage

\bibliography{references}

\bigskip

\end{document}